\documentclass[a4paper,12pt]{article}

 \usepackage{amsfonts}
\usepackage{verbatim}

\begin{document}


\vskip 0.9cm

\begin{center}

{\Large \sf Symmetry invariance, anticommutativity and nilpotency in 
BRST approach to QED: superfield formalism}\\

\vskip 2cm

{\sf R. P. Malik}\\  
{\it Physics Department, Centre of
Advanced Studies,}\\
{\it Banaras
Hindu University, Varanasi-221 005 (U.P.), India}\\

and\\

{\it DST Centre for Interdisciplinary Mathematical Sciences,}\\
{\it Banaras
Hindu University, Varanasi-221 005 (U.P.), India}\\

and\\
 
{\it SISSA, Strada Costiera, Trieste, Italy}\\ 

{\small E-mails: rudra.prakash@hotmail.com; malik@bhu.ac.in}\\

\end{center}

\vskip 1.5cm

\noindent
{\sf Abstract:}
We provide the geometrical interpretation for the Becchi-Rouet-Stora-Tyutin
(BRST) and anti-BRST symmetry invariance of the Lagrangian density of
a four (3 + 1)-dimensional (4D) interacting $U(1)$ gauge theory within the
framework of superfield approach to BRST formalism. This interacting theory, 
where  there is an explicit coupling between the $U(1)$ gauge field and matter 
(Dirac) fields, is considered on a (4, 2)-dimensional supermanifold
parametrized by the four spacetime variables $x^\mu (\mu = 0, 1, 2, 3)$ and
a pair of Grassmannian variables $\theta$ and $\bar\theta$ (with $\theta^2 
= \bar \theta^2 = 0, \theta \bar\theta + \bar \theta \theta = 0$). We express 
the Lagrangian density and (anti-)BRST charges in the language of the superfields 
and show that (i) the (anti-)BRST invariance of the 4D Lagrangian density is
equivalent to the translation of the super Lagrangian density along the Grassmannian
direction(s) ($\theta$ and/or $\bar\theta$) of the (4, 2)-dimensional supermanifold such that the
outcome of the above translation(s) is zero, and (ii) the anticommutativity and nilpotency
of the (anti-)BRST charges are the automatic consequences of our superfield formulation.\\

\noindent
PACS numbers: 11.15.-q, 12.20.-m, 03.70.+k\\

\noindent
{\it Keywords:} QED with Dirac fields, nilpotent (anti-)BRST symmetry invariance,
superfield formalism, horizontality condition, gauge invariant condition on
the matter superfields, geometrical interpretation(s)\\

\newpage

\section{Introduction}

The usual superfield approach [1-8] to BRST formalism has been
very successfully applied to the case of  4D (non-)Abelian
1-form ($A^{(1)} = dx^\mu A_\mu$) gauge theories. In this
approach, one constructs a super curvature 2-form $\tilde F^{(2)}
= \tilde d \tilde A^{(1)} + i \tilde A^{(1)} \wedge \tilde
A^{(1)}$ by exploiting the Maurer-Cartan equation in the language
of the super 1-form gauge connection $\tilde A^{(1)}$ and the
super exterior derivative $\tilde d = d Z^M \partial_M \equiv
dx^\mu \partial_\mu +
d\theta \partial_\theta + d \bar\theta \partial_{\bar\theta}$
(with $\tilde d^2 = 0$) that are defined on a (4, 2)-dimensional
supermanifold, parametrized by the superspace variables $Z^M =
(x^\mu, \theta, \bar\theta)$ (with the super derivatives $\partial_M 
= (\partial_\mu, \partial_\theta, \partial_{\bar\theta})$).

The above super curvature is subsequently equated to the
ordinary 2-form curvature $F^{(2)} = d A^{(1)} + i A^{(1)} \wedge
A^{(1)}$ (with $d = dx^\mu \partial_\mu$ and $A^{(1)} = dx^\mu A_\mu$) 
defined on the 4D ordinary flat Minkowskian spacetime
manifold that is parametrized by the ordinary spacetime variable
$x^\mu (\mu = 0, 1, 2, 3)$. This restriction, popularly known as
the horizontality condition, leads to the derivation of the
nilpotent (anti-)BRST symmetry transformations for the gauge and
(anti-)ghost fields of the 4D (non-)Abelian 1-form gauge theories.

The key reasons behind the emergence of the nilpotent (anti-)BRST
symmetry transformations for the gauge and (anti-)ghost fields,
due to the above horizontality condition\footnote{This condition
has been christened as the soul-flatness condition in [9].} (HC),
are

(i) the nilpotency of the (super) exterior derivatives $(\tilde d)
d$ which play very important roles in the above HC, and

(ii) the super 1-form connection $\tilde A^{(1)} = dZ^M \tilde
A_M$ involves the vector superfield $\tilde A_M$ that consists of
the multiplet superfields (${\cal B}_\mu, {\cal F}, \bar {\cal
F}$) which are nothing but the generalizations of the gauge and (anti-)ghost
fields $(A_\mu, C, \bar C)$. The latter are the basic fields of the 4D
(non-)Abelian 1-form gauge theories.

The above equality (i.e. $\tilde F^{(2)} = F^{(2)}$),
due to the HC\footnote{It is clear that, for the Abelian 1-form theory,
we have $\tilde F^{(2)} = \tilde d \tilde A^{(1)}$ and $F^{(2)} = d A^{(1)}$.}, 
implies that the ordinary curvature 2-form
$F^{(2)}$ remains unaffected by the presence of the Grassmannian
coordinates $\theta$ and $\bar\theta$ (with $\theta^2 =
\bar\theta^2 = 0, \theta \bar\theta + \bar \theta \theta = 0$) of
the superspace variable $Z^M$. This restriction, however,
does not shed any light on the derivation of the nilpotent
(anti-)BRST symmetry transformations that are associated with the
matter (e.g. Dirac, complex scalar, etc.) fields of an interacting
4D (non-)Abelian 1-form gauge theory.

In a recent set of papers [10-14], the above HC has been extended
so as to derive the nilpotent (anti-)BRST symmetry transformations
for the matter (or analogous) fields within the framework of the
superfield approach to BRST formalism without spoiling the cute
geometrical interpretations for the nilpotent (anti-)BRST symmetry
transformations (and their corresponding generators) that emerge
from the application of the HC {\it alone}.

In fact, in a set of a couple of
papers [13-14], we have been able to generalize the HC by a gauge
invariant restriction (GIR) on the matter superfields (defined on
the above (4, 2)-dimensional supermanifold) which enables us to
derive the nilpotent (anti-)BRST symmetry transformations {\it
together} for the gauge, (anti-)ghost and matter fields of a 4D
interacting (non-)Abelian gauge theory in one stroke. In this {\it
single} GIR on the matter superfields of the above supermanifold,
the (super) covariant derivatives and their intimate connection
with the (super) curvature 2-forms, play a very decisive role.

In the earlier works on the superfield formulation [1-14], the
nilpotent (anti-)BRST symmetry invariance of the physical 4D
Lagrangian density of the (non-)Abelian 1-form gauge theories has
{\it not yet} been captured. In our previous endeavours [15-17], we have
attempted to capture the nilpotent symmetry invariance of the 2D
(non-)Abelian 1-form gauge theories (without any interaction with
matter fields) within the framework of the superfield approach to
BRST formalism. However, these theories are found to be
topological in nature and they are endowed with the nilpotent
(anti-)BRST as well as nilpotent {\it (anti-)co-BRST} symmetry
transformations.

In our very recent
paper [18], we have been able to provide the geometrical
interpretation for the (anti-)BRST invariance of the 4D free
(non-)Abelian 1-form gauge theories (where there is no interaction
with the matter fields) within the framework of the superfield
formalism. In this work [18], we have also provided the reasons
behind the uniqueness of the above symmetry transformations 
(and their invariances) and
furnished the logical arguments for the non-existence of the
on-shell nilpotent (anti-)BRST symmetry transformations {\it together} for the 4D
non-Abelian 1-form gauge theory.

The central theme of our present paper is to generalize the key
results of our earlier work in [18] to the more general case of
an interacting $U(1)$ gauge theory where there is an explicit
coupling between the 1-form U(1) gauge field and the Noether
conserved current constructed with the matter (Dirac) fields. We
find that the GIR on the matter (Dirac) superfields (defined on
the above (4, 2)-dimensional supermanifold) enables us to derive
the exact nilpotent (anti-)BRST symmetry transformations for the
matter (Dirac) fields which can {\it never} be obtained by
exploiting the HC {\it alone}. The above GIR also provides a
meeting-ground for the HC of the usual superfield formalism [1-9]
and a gauge {\it invariant} condition on the matter superfields.

For our central objective of encapsulating the (anti-)BRST invariance
of the 4D Lagrangian density of the interacting U(1) gauge theory
within the framework of the superfield approach to BRST formalism,
the following key points are of utmost importance, namely;

(i) the application of the HC enables us to demonstrate that the kinetic
energy term of the U(1) gauge field remains independent of the Grassmannian
variables when it is expressed in terms of the gauge superfields
(that are obtained after the application of the HC), and

(ii)  the application of the above GIR on the matter superfields
enables us to show that all the terms containing the matter
(Dirac) superfields are independent of the Grassmannian variables
when they are expressed in terms of the superfields that are
obtained after the application of the HC as well as the GIR on the
matter superfields (see, e.g. (23) below).

The above key restrictions (i.e. HC and GIR) enable us to express
the total Lagrangian density of the 4D interacting U(1) gauge
theory with Dirac fields in the language of the superfields, in
such a manner that, ultimately, a partial derivative w.r.t.
$\theta$ and/or $\bar\theta$ on it becomes zero. In other words,
the total super Lagrangian density (defined on the (4,
2)-dimensional supermanifold) becomes independent of
the Grassmannian variables.

The above observation, in
turn, implies that the corresponding 4D Lagrangian density of the
parent theory (defined on the 4D ordinary spacetime manifold)
becomes automatically (anti-)BRST invariant. Stated in the
language of geometry on the above  supermanifold, the
translation of the above super Lagrangian density (defined in
terms of the superfields obtained after the application of the HC
and GIR) along either of the Grassmannian directions (i.e. $\theta$
and/or $\bar\theta$) of the above  supermanifold
becomes zero. This result is consistent with our earlier observation in
[18].

The main motivating factors that have contributed to our curiosity
to carry out the present investigation are as follows. 
First, it is very important for us to generalize our earlier
work [18] to the case where there is an explicit coupling
between the U(1) gauge field and matter (Dirac) fields. 
Second, it
is a challenge to check the validity of the geometrical
interpretations, provided for the (anti-)BRST invariance in our
earlier work [18], to the present {\it interacting} case. Third, we explicitly
express the (anti-)BRST charges in terms of the superfields
and prove their nilpotency and anticommutativity properties.
Finally, our present attempt is a 
modest step towards our main goal of
applying the superfield formalism to the case of higher p-form ($ p \geq 2$)
gauge theories which have become important in the context of string theories.

Our present paper is organized as follows.

In section 2, we give a brief synopsis of the nilpotent (anti-)BRST
symmetry invariance of the Lagrangian density of a 4D interacting $U(1)$
gauge theory where there is an explicit coupling between the U(1)
gauge field and the Noether conserved current, constructed with the
help of the Dirac fields.

We exploit, in section 3, the horizontality condition (HC) to 
express the kinetic energy, gauge-fixing, and ghost terms
in the language of the superfields (derived after the application of HC).

Section 4 deals with a GIR on the matter superfields to obtain the (anti-)BRST symmetry
transformations for the matter fields and to express the kinetic
term, interaction term and mass term of the Dirac fields in
the language of the superfields, obtained after the application of
the HC and GIR.

In section 5,  we express the (anti-)BRST charges in the language of
the superfields obtained after the application of the HC and GIR. This
exercise enables us to prove the nilpotency and anticommutativity
of the above charges in a simple manner from which the geometrical meanings
ensue.

Finally, in section 6, we make some
concluding remarks and point out a few future directions for
further investigations.

Our Appendix A is devoted to the derivation of the nilpotent
(anti-)BRST symmetry transformations {\it together} for the gauge,
matter and (anti-)ghost fields of the theory from a {\it single} GIR on the
matter superfields.

\section{Nilpotent (anti-)BRST symmetry invariance in QED: Lagrangian formalism}

\noindent We begin with the following nilpotent (anti-)BRST
symmetry invariant Lagrangian density of the 4D interacting
Abelian 1-form U(1) gauge theory in the Feynman gauge \footnote{We
follow here the convention and notations such that the flat 4D
Minkowski spacetime manifold is characterized by a metric
$\eta_{\mu\nu}$ with the signatures $(+ 1,
- 1, -1, -1)$ so that $\Box = \eta_{\mu\nu} \partial^\mu
\partial^\nu = (\partial_0)^2 - (\partial_i)^2$ and $P \cdot Q =
\eta_{\mu\nu} P^\mu Q^\nu\equiv P_0 Q_0 - P_i Q_i$ is the dot
product between two non-null 4-vectors $P_\mu$ and $Q_\mu$.
The Greek indices $\mu, \nu,
\kappa,.....= 0 , 1, 2, 3$ correspond to the spacetime directions
and the Latin indices $i, j, k, .... = 1, 2, 3$ stand only for the
space directions on the above 4D flat Minkowskian
spacetime manifold.} (see,
e.g. [9])
\begin{eqnarray}
{\cal L}_b &=&
- \frac{1}{4} F^{\mu\nu} F_{\mu\nu} + \bar \psi (i \gamma^\mu D_\mu - m)
\psi + B (\partial \cdot A) + \frac{1}{2} B^2
- i \partial_\mu \bar C \partial^\mu C  \nonumber\\
&\equiv& {\cal L}^{(g)}_b + {\cal L}^{(d)}_b.
\end{eqnarray}
In the above, the kinetic energy term for the 1-form gauge field is
constructed with the help of the curvature tensor $F_{\mu\nu}$
which is derived from the 2-form $F^{(2)} = (1/2!) (dx^\mu \wedge
dx^\nu) F_{\mu\nu}$. The latter emerges (i.e. $F^{(2)} = d
A^{(1)}$) when the exterior derivative $d = dx^\mu \partial_\mu$
(with $d^2 = 0$) acts on the 1-form connection $A^{(1)} = dx^\mu
A_\mu$ that defines the gauge potential $A_{\mu}$ of our present
theory. The Nakanishi-Lautrup auxiliary scalar field $B$ is
invoked to linearize the gauge-fixing term $[- (1/2) (\partial
\cdot A)^2]$. The latter requires, for the nilpotent (anti-)BRST
symmetry invariance in the theory, the fermionic (i.e. $C^2 = \bar
C^2 = 0, C \bar C + \bar C C = 0$) (anti-)ghost fields $(\bar C)C$
which play a central role in the proof of the unitarity of the
theory. The covariant derivative $D_\mu \psi = \partial_\mu \psi +
i e A_\mu \psi$ generates the interaction term (i.e. $-e \bar\psi
\gamma^\mu A_\mu \psi$) between the gauge field $A_\mu$ and Dirac
fields $\psi$ and $\bar\psi$ (with mass $m$ and charge $e$). Here
$\gamma^\mu$'s are the standard hermitian $4 \times 4$ Dirac matrices
in 4D.

The above Lagrangian (1) has been split into two parts ${\cal
L}^{(g)}_b$ and ${\cal L}^{(d)}_b$ for our later convenience. The
former corresponds to the kinetic term, gauge-fixing term
and Faddeev-Popov ghost term for the 1-form gauge and fermionic
(anti-)ghost fields of the theory and the latter corresponds to
all the terms in the Lagrangian density that necessarily possess
the Dirac fields.

The following off-shell nilpotent $(s_{(a)b}^2 = 0$) and
anticommuting ($s_b s_{ab} + s_{ab} s_b = 0$) infinitesimal
(anti-)BRST symmetry transformations $s_{(a)b}$\footnote{We follow
here the notation and convention adopted in [10-14]. In their
totality, the (anti-)BRST symmetry transformations $\delta_{(A)B}$
are the product (i.e. $\delta_{(A)B} = \eta s_{(a)b}$) of an
anticommuting ($\eta C = = C \eta, \eta \psi = - \psi \eta$ etc.)
spacetime independent parameter $\eta$ and $s_{(a)b}$ such that
the operator form of the nilpotency of $\delta_{(A)B}$ is traded
with that of $s_{(a)b}$.}
\begin{eqnarray}
&& s_b A_\mu = \partial_\mu C, \qquad s_b  C = 0, \qquad s_b \bar C
= i B, \qquad s_b B = 0, \nonumber\\ && s_b \psi = - i e C \psi,\;
\qquad s_b \bar \psi = - i e \bar \psi C,\; \qquad s_b F_{\mu\nu} =
0,
\end{eqnarray}
\begin{eqnarray}
&& s_{ab} A_\mu = \partial_\mu \bar C, \qquad s_{ab} \bar C = 0,
\qquad s_{ab} C = - i B, \qquad s_{ab} B = 0, \nonumber\\ && s_{ab}
\psi = - i e \bar C \psi, \;\qquad s_{ab} \bar \psi = - i e \bar
\psi \bar C, \qquad \; s_{ab} F_{\mu\nu} = 0,
\end{eqnarray}
leave the above Lagrangian density (1) quasi-invariant because it
transforms to a total spacetime derivative under the above transformations
(cf. (26) below).
The key features of the above transformations are:

(i) the curvature tensor, owing its origin to the cohomological
operator $d = dx^\mu \partial_\mu$ (with $d^2 = 0$), remains
invariant under both the above nilpotent symmetry transformations
(i.e. $s_{(a)b} F_{\mu\nu} = 0$),

(ii) the total terms involving the Dirac fields (i.e. $\bar \psi
(i \gamma^\mu D_\mu - m) \psi$) also remain invariant under
$s_{(a)b}$ (i.e. $s_{(a)b} [\bar\psi (i \gamma^\mu D_\mu - m)
\psi] = 0$),

(iii) the nilpotency (i.e. $d^2 = 0$) of the exterior derivative
$d = dx^\mu \partial_\mu$ is replicated in the language of the
nilpotency of the above symmetry transformations $s_{(a)b}$ (i.e.
$s_{(a)b}^2 = 0$), and

(iv) there is a deep connection between the exterior derivative
and the above nilpotent transformations which will be exploited in
the superfield approach to BRST formalism (see, sections 3,4 and Appendix
below).

It can be checked that the gauge-fixing and Faddeev-Popov ghost
terms of the Lagrangian density (1) can be written, in the exact
form(s), as
\begin{eqnarray}
- s_b \Bigl [i \bar C \{(\partial \cdot A) + \frac{1}{2} B \} \Bigr ],
s_{ab} \Bigl [i  C \{(\partial \cdot A) + \frac{1}{2} B \} \Bigr ],
s_b s_{ab} \Bigl ( \frac{i}{2} A_\mu A^\mu + \frac{1}{2} C \bar C \Bigr ),
\end{eqnarray}
modulo some total spacetime derivative terms which do not affect the dynamics of the theory.
The above expressions provide a simple proof for the nilpotent symmetry invariance
of the Lagrangian density (1) because of

(i) the nilpotency of the transformations
$s_{(a)b}$ (i.e. $s_{(a)b}^2 = 0$),

(ii) the invariance of the curvature term (i.e. $s_{(a)b}
F_{\mu\nu} = 0$), and

(iii) the invariance of the terms involving Dirac fields (i.e.
$s_{(a)b}\; [\bar \psi (i \gamma^\mu D_\mu - m) \psi] = 0$)
under the nilpotent (anti-)BRST symmetry transformations $s_{(a)b}$.

\section{Symmetry transformations for the gauge and ghost fields: horizontality condition}

We exploit here the usual superfield approach [1-9] to obtain the
nilpotent (anti-)BRST symmetry transformations for the gauge and
(anti-)ghost fields of the Lagrangian density (1). To this end in
mind, first of all, we generalize the 4D local fields $(A_\mu (x),
C (x), \bar C (x))$ to the superfields $({\cal B}_\mu
(x,\theta,\bar\theta)$, ${\cal F} (x, \theta, \bar\theta)$, $\bar
{\cal F} (x,\theta,\bar\theta))$ that are defined on a (4,
2)-dimensional supermanifold, parametrized by the superspace
variables $Z^M = (x^\mu, \theta, \bar\theta)$. In terms of these
superfields, we can define a super 1-form connection as
\begin{equation}
\tilde A^{(1)} = d Z^M\; \tilde A_M
= dx^\mu \;{\cal B}_\mu + d \theta\;
\bar {\cal F} + d \bar\theta \;{\cal F},
\end{equation}
where $\tilde A_M$ is the vector superfield with the component
multiplet fields as $({\cal B}_\mu, {\cal F}, \bar {\cal F})$ and
$ dZ^M = (dx^\mu, d \theta, d \bar\theta)$ is the superspace
differential.

The above component superfields can be expanded in terms of the basic fields
$(A_\mu, C, \bar C)$, auxiliary field $B$ and
secondary fields as (see, e.g. [4-7])
\begin{eqnarray}
{\cal B}_\mu (x, \theta, \bar\theta) &=& A_\mu (x) + \theta \;\bar
R_\mu (x) + \bar\theta \;R_\mu (x) + i\; \theta \;\bar\theta\;
S_\mu (x), \nonumber\\ {\cal F} (x, \theta, \bar\theta) &=&  C (x)
+ i\; \theta \;\bar B (x) + i\; \bar\theta \; {\cal B} (x) + i
\;\theta\; \bar\theta\; s (x), \nonumber\\ \bar {\cal F} (x,
\theta, \bar\theta) &=& \bar C (x) + i \;\theta \;\bar {\cal B}
(x) + i \;\bar\theta\; B (x) + i \;\theta \;\bar\theta\; \bar s
(x),
\end{eqnarray}
where the secondary fields are $\bar B (x), {\cal B} (x), \bar
{\cal B} (x), s(x), \bar s(x)$. It will be noted that, in the
limit $(\theta, \bar\theta) \to 0$, we retrieve our basic fields
$(A_\mu, C, \bar C)$. In the above expansion, the bosonic and
fermionic component fields do match with each-other. The exact
expressions for the secondary fields can be derived in terms of
the basic fields of the theory if we exploit the celebrated HC.

Let us recall our observation after the (anti-)BRST symmetry
transformations (2) and (3). These transformations $s_{(a)b}$ owe
their origin to the exterior derivative $d = dx^\mu \partial_\mu$
which plays a very important role in the application of the HC. To
this end in mind, let us generalize the 4D ordinary $d$ to its
counterpart on the (4, 2)-dimensional supermanifold, as
\begin{equation}
d \to \tilde d = d Z^M \partial_M
= dx^\mu \partial_\mu + d \theta \partial_\theta
+ d \bar\theta \partial_{\bar\theta}, \; \qquad
\partial_M = (\partial_\mu, \partial_\theta, \partial_{\bar\theta}).
\end{equation}
The HC is the requirement that the super 2-form $\tilde F^{(2)} =
\tilde d \tilde A^{(1)}$, defined on the (4, 2)-dimensional
supermanifold, should be equal (i.e. $\tilde F^{(2)} = F^{(2)}$)
to the ordinary 4D 2-form $F^{(2)} = d A^{(1)}$. Physically, this
amounts to the restriction that the gauge (i.e. (anti-)BRST)
invariant quantities, which are the components of
the curvature tensor $F_{\mu\nu}$, should remain {\it unaffected}
by the presence of the Grassmannian variables $\theta$ and
$\bar\theta$.

The explicit computations for $\tilde F^{(2)}$ (from (5) and (7))
and its subsequent equality with the ordinary 4D 2-form $F^{(2)}$,
leads to the following relationships between the secondary fields
and basic fields of the theory [10-14]
\begin{eqnarray}
&& R_\mu = \partial_\mu C, \qquad \bar R_\mu = \partial_\mu \bar C,
\qquad S_\mu = \partial_\mu B \equiv - \partial_\mu \bar B,
\nonumber\\ && B + \bar B = 0, \qquad {\cal B} = 0, \qquad \bar
{\cal B} = 0, \qquad s = 0, \qquad \bar s = 0.
\end{eqnarray}
Insertion of these values into the expansion (6) of the
superfields leads to the following final expansion
\begin{eqnarray}
{\cal B}^{(h)}_\mu (x, \theta, \bar\theta)
&=& A_\mu (x) + \theta \;\partial_\mu
\bar C (x) + \bar\theta \; \partial_\mu C (x) + i\; \theta \;\bar\theta\;
\partial_\mu B (x) \nonumber\\
&\equiv& A_\mu (x) + \theta  (s_{ab} A_\mu (x)) + \bar \theta (s_b
A_\mu (x)) + \theta  \bar\theta\; (s_b s_{ab} A_\mu (x)),
\nonumber\\ {\cal F}^{(h)} (x, \theta, \bar\theta) &=&  C (x) -
i\; \theta \; B (x) \nonumber\\ &\equiv& C (x) + \theta\; (s_{ab}
C (x)) + \bar\theta\; (s_b C (x)) + \theta \bar\theta (s_b s_{ab}
C (x)), \nonumber\\ \bar {\cal F}^{(h)} (x, \theta, \bar\theta)
&=& \bar C (x) + i \;\bar\theta\; B (x) \nonumber\\ &\equiv& \bar C (x)
+ \theta\; (s_{ab} \bar C (x)) + \bar\theta\; (s_b \bar C (x)) + \theta
\bar\theta (s_b s_{ab} \bar C (x)),
\end{eqnarray}
where the following points are important, namely;

(i) the superscript $(h)$ on the superfields corresponds to the superfields
obtained after the application of the HC,

(ii) the transformations $s_b C = 0$ and $s_{ab} \bar C = 0$ have
been taken into account in the above uniform expansions,

(iii) the super curvature tensor $\tilde F^{(h)}_{\mu\nu} =
\partial_\mu {\cal B}^{(h)}_\nu - \partial_\nu {\cal B}^{(h)}_\mu$
is found to be equal to the ordinary curvature tensor $F_{\mu\nu}
=
\partial_\mu A_\nu - \partial_\nu A_\mu$ because all the
Grassmannian dependent terms cancel out to become zero, and

(iv) there is a very intimate relationship between the (anti-)BRST
symmetry transformations acting on a 4D field and the
translational generators (along the Grassmannian directions of the
supermanifold) acting on the corresponding (4, 2)-dimensional
superfield obtained after the application HC. Mathematically,
this statement can be expressed as
\begin{equation}
\mbox{Lim}_{\theta \to 0} \frac{\partial}{\partial \bar\theta}
\tilde \Omega^{(h)} (x, \theta, \bar\theta) = s_b \Omega (x), \quad
\mbox{Lim}_{\bar \theta \to 0} \frac{\partial}{\partial\theta}
\tilde \Omega^{(h)} (x, \theta, \bar\theta) = s_{ab} \Omega (x),
\end{equation}
where $\Omega (x)$ is the ordinary 4D generic local field and
$\tilde \Omega^{(h)} (x,\theta,\bar\theta)$ is the corresponding
generic superfield derived after the application of the HC.

The above superfields would be used to express the kinetic 
term for the U(1) gauge field, gauge-fixing term for the photon
field and Faddeev-Popov ghost terms for the (anti-)ghost fields of
the theory. These can be expressed in the following {\it three}
different and distinct forms (see, e.g. [18])
\begin{eqnarray}
\tilde {\cal L}^{(1)}_{(g)} &=& - \frac{1}{4} \tilde
F^{(h)}_{\mu\nu} \tilde F^{\mu\nu (h)} + \mbox{Lim}_{\theta \to 0}
\frac{\partial}{\partial \bar\theta} \Bigl [ - i\; \bar {\cal
F}^{(h)} \bigl ( \partial^\mu {\cal B}^{(h)}_\mu + \frac{1}{2} \;
B \bigr )\;\Bigr ], \nonumber\\ \tilde {\cal L}^{(2)}_{(g)} &=& -
\frac{1}{4} \tilde F^{(h)}_{\mu\nu} \tilde F^{\mu\nu (h)} +
\mbox{Lim}_{\bar\theta \to 0} \frac{\partial}{\partial \theta}
\Bigl [ + i\;  {\cal F}^{(h)} \bigl ( \partial^\mu {\cal
B}^{(h)}_\mu + \frac{1}{2} \; B \bigr )\; \Bigr ], \nonumber\\
 \tilde {\cal L}^{(3)}_{(g)} &=& - \frac{1}{4} \tilde F^{(h)}_{\mu\nu} \tilde
F^{\mu\nu (h)} + \frac{\partial}{\partial \bar\theta}\;
\frac{\partial}{\partial\theta} \Bigl [ + \frac{i}{2} {\cal
B}^{\mu (h)} {\cal B}^{(h)}_\mu + \frac{1}{2} \; {\cal F}^{(h)}
\bar {\cal F}^{(h)} \Bigr ],
\end{eqnarray}
where the subscript (g) on the above super Lagrangians stands for
the terms in the 4D Lagrangian density that correspond to the
gauge and (anti-)ghost fields. In other words, we have encapsulated
the kinetic term, gauge-fixing term and Faddeev-Popov ghost
term of the 4D Lagrangian density (1) in the language of the
superfields derived after the application of the HC.

It is very interesting to note that the (anti-)BRST invariance of
the kinetic term, gauge-fixing term and Faddeev-Popov ghost
term can be captured in the language of the translations of the
above super Lagrangian densities $\tilde {\cal L}^{(1,2,3)}_{(g)}$
along the Grassmannian directions. Mathematically, the nilpotent
BRST invariance can be expressed as follows
\begin{equation}
s_b {\cal L}^{(g)}_b = 0 \Leftrightarrow \mbox{Lim}_{\theta \to 0}
\frac{\partial} {\partial\bar\theta} \tilde {\cal L}^{(1)}_{(g)} =
0, \qquad s_b^2 = 0 \Leftrightarrow (\mbox{Lim}_{\theta \to 0}
\frac{\partial} {\partial\bar\theta})^2 = 0,
\end{equation}
\begin{equation}
s_b {\cal L}^{(g)}_b = 0 \Leftrightarrow \frac{\partial}
{\partial\bar\theta} \tilde {\cal L}^{(3)}_{(g)} = 0, \qquad s_b^2
= 0 \Leftrightarrow ( \frac{\partial} {\partial\bar\theta})^2 = 0.
\end{equation}
Similarly, the anti-BRST invariance of the kinetic,
gauge-fixing and ghost terms can be expressed, in the language of
the superfields, as follows
\begin{equation}
s_{ab} {\cal L}^{(g)}_b = 0 \Leftrightarrow \mbox{Lim}_{\bar
\theta \to 0} \frac{\partial} {\partial\theta} \tilde {\cal
L}^{(2)}_{(g)} = 0, \qquad s_{ab}^2 = 0 \Leftrightarrow
(\mbox{Lim}_{\bar\theta \to 0} \frac{\partial} {\partial\theta})^2
= 0,
\end{equation}
\begin{equation}
s_{ab} {\cal L}^{(g)}_b = 0 \Leftrightarrow \frac{\partial}
{\partial\theta} \tilde {\cal L}^{(3)}_{(g)} = 0, \qquad s_{ab}^2 =
0 \Leftrightarrow ( \frac{\partial} {\partial\theta})^2 = 0.
\end{equation}
Thus, we note that the Grassmannian independence of the super
Lagrangian densities, defined in terms of the superfields on the (4, 2)-dimensional
supermanifold, automatically implies the (anti-)BRST invariance of
the 4D Lagrangian density defined in terms of 4D local fields
taking their values on a 4D flat Minkowskian spacetime manifold.

\section{(Anti-)BRST symmetry transformations for
Dirac fields: gauge invariant condition}

Unlike the superfields ${\cal B}_\mu (x,\theta,\bar\theta), {\cal
F} (x,\theta,\bar\theta), \bar {\cal F} (x,\theta,\bar\theta)$
that form a vector supermultiplet (cf. previous section), the
fermionic matter superfields $\Psi (x, \theta, \bar\theta)$ and
$\bar \Psi (x,\theta,\bar\theta)$ (which are the generalizations
of the 4D matter Dirac fields $\psi (x)$ and $\bar\psi (x)$ of the
Lagrangian density (1) onto the (4, 2)-dimensional supermanifold) do
not belong to any supermultiplet. Thus, it appears that there is
no connection between the superfields $({\cal B}_\mu, {\cal F},
\bar {\cal F})$  and the matter superfields ($\Psi (x, \theta,
\bar\theta), \bar \Psi (x,\theta,\bar\theta)$). However, there is
one {\it gauge invariant} relationship in which the matter
superfields do `talk' with the super gauge connection  $\tilde
A^{(1)}$ of (5). We exploit this relationship and impose the
following GIR on the matter superfields 
\begin{equation}
\bar \Psi (x, \theta, \bar\theta) \;
\bigl [ \tilde d + i e \tilde A^{(1)}_{(h)}
\bigr ]\; \Psi (x, \theta, \bar\theta)
= \bar \psi (x) (d + i e A^{(1)}) \psi (x).
\end{equation}
The r.h.s. of the above equation is
a U(1) gauge (i.e. (anti-)BRST) invariant quantity because it is
connected with the covariant derivative.

The relationship (16) is interesting on the following grounds.
First, it is a gauge (i.e. (anti-)BRST) invariant quantity. Thus,
it is physical in some sense. Second, it will be noted that, on
the l.h.s. of (16), we have $\tilde A^{(1)}_{(h)}$ which is
derived after the application of HC\footnote{The HC is basically a
gauge {\it covariant} restriction for the discussion of the
non-Abelian gauge theory. It, however, reduces to a GIR for the
Abelian U(1) gauge theory.}. Thus, HC of the previous section and
GIR of our present section (cf. (16)) are intimately connected. In
fact, the explicit form of $\tilde A^{(1)}_{(h)} = dx^\mu \;{\cal
B}^{(h)}_\mu + d \theta\; \bar {\cal F}^{(h)} + d \bar \theta
\;{\cal F}^{(h)}$ is such that the whole expansion of (9) is going
to play a very decisive role in the determination of the exact
nilpotent (anti-)BRST symmetry transformations for the matter
fields in the language of the (anti-)ghost fields and matter
fields themselves.

To find out the impact of the above restriction, we have to expand
the matter superfields along the Grassmannian $\theta$ and
$\bar\theta$ directions of the (4, 2)-dimensional supermanifold as
follows
\begin{eqnarray}
\Psi (x, \theta, \bar\theta) &=& \psi (x) + i \; \theta\; \bar b_1
(x) + i\; \bar\theta\; b_2(x) + i \theta\; \bar\theta\; f(x),
\nonumber\\ \bar \Psi (x, \theta, \bar\theta) &=& \bar \psi (x) +
i \; \theta\; \bar b_2 (x) + i\; \bar\theta\; b_1(x) + i \theta\;
\bar\theta\; \bar f(x),
\end{eqnarray}
where $\psi (x)$ and $\bar\psi (x)$ are the 4D basic Dirac fields
of the Lagrangian density (1) and $b_1, \bar b_1, b_2, \bar b_2,
f, \bar f$ are the secondary fields which will be expressed in
terms of the basic fields of the Lagrangian density (1) due to the
above GIR (16). In the limit $(\theta, \bar\theta) \to 0$, we
retrieve our basic 4D local fields $\psi$ and $\bar\psi$ and
bosonic $(b_1, b_2, \bar b_1, \bar b_2)$ and fermionic $(\psi,
\bar \psi, f, \bar f$) degrees of freedom do match in the above
expansion. This is consistent with the basic requirements of a
true supersymmetric field theory.

The explicit computation of the l.h.s. of GIR (16) and its subsequent comparison
with the r.h.s., yields the following relationship between the secondary fields
and the basic fields of the theory (see [13] for details)
\begin{eqnarray}
b_1 &=& - e \bar \psi C,\; \qquad b_2 = - e C \psi,\; \qquad \bar b_1 = - e \bar C \psi,\;
\qquad \bar b_2 = - e \bar\psi \bar C, \nonumber\\
f &=& - i e [ B + e \bar C C ] \psi,\; \qquad \;\; \bar f = + i e \bar \psi\;
[ B + e C \bar C ].
\end{eqnarray}
The insertions of the above values into the expansion of the matter
superfields (17), lead to the following explicit expansions
\begin{eqnarray}
\Psi^{(G)} (x, \theta, \bar\theta) &=& \psi (x) + \theta \; (-i e \bar C \psi (x))
+ \bar\theta\; (- i e C \psi (x)) \nonumber\\ &+& \theta \;\bar\theta\;
[ e (B + e \bar C C) \psi (x) ] \nonumber\\
&\equiv& \psi (x) + \theta\; (s_{ab} \psi (x)) + \bar\theta\; (s_b \psi (x))
+ \theta\; \bar\theta\; (s_b s_{ab} \psi (x)), \nonumber\\
\bar \Psi^{(G)} (x, \theta, \bar\theta) &=& \bar \psi (x) + \theta \; (-ie
\bar \psi (x) \bar C)
+ \bar\theta\; (- i e \bar \psi (x) C) \nonumber\\  &+& \theta \;\bar\theta\;
[- e \bar \psi (x) (B + e C \bar C) ] \nonumber\\
&\equiv& \bar \psi (x) + \theta\; (s_{ab} \bar\psi (x))
+ \bar\theta\; (s_b \bar\psi (x))
+ \theta\; \bar\theta\; (s_b s_{ab} \bar\psi (x)).
\end{eqnarray}
The superscript $(G)$ on the above superfields denotes the fact
that they have been obtained after the application of the GIR (cf.
equation (16)).

It is very interesting to note, at this stage, that the GIR (cf. (16))
on the matter superfields leads to

(i) the exact and unique derivation of the nilpotent (anti-)BRST
symmetry transformations for the matter fields $\psi (x) $ and
$\bar\psi (x)$, and

(ii) the geometrical interpretation for the (anti-)BRST symmetry
transformations as the translational generators along the
Grassmannian directions $\theta$ and $\bar\theta$ of the above
supermanifold. As a consequence, we obtain the analogue of the
equation (10), as
\begin{equation}
\mbox{Lim}_{\theta \to 0} \frac{\partial}{\partial \bar\theta}
\tilde \Omega^{(G)} (x, \theta, \bar\theta) = s_b \Omega (x), \quad
\mbox{Lim}_{\bar \theta \to 0} \frac{\partial}{\partial\theta}
\tilde \Omega^{(G)} (x, \theta, \bar\theta) = s_{ab} \Omega (x),
\end{equation}
where $\tilde \Omega^{(G)} (x, \theta, \bar\theta)$ stands for the expansions
(19) for the matter superfields, obtained after the application of the GIR  (16).
The generic field $\Omega (x)$ stands for the 4D matter Dirac fields $\psi (x)$ and $\bar\psi (x)$.

 There is an interesting consequence
due to our expansion in (19). It is straightforward to check that
the following equation
\begin{equation}
\bar \Psi^{(G)} (x, \theta, \bar\theta) \; \Psi^{(G)} (x, \theta, \bar\theta) =
\bar \psi (x) \; \psi (x),
\end{equation}
is automatically satisfied. This observation implies, ultimately,
that the equation (16) can be re-expressed as
\begin{equation}
\bar \Psi^{(G)} (x, \theta, \bar\theta) \;\bigl [ \tilde d + i e
\tilde A^{(1)}_{(h)} - m
\bigr ]\; \Psi^{(G)} (x, \theta, \bar\theta) = \bar \psi (x) (d + i e A^{(1)} - m) \psi (x).
\end{equation}
Physically, the above restriction implies that the total Dirac
part of the Lagrangian density (1) (i.e. ${\cal L}^{(d)}$) remains
unaffected due to the presence of the Grassmannian variables on
the (4, 2)-dimensional supermanifold, on which, our 4D interacting
$U(1)$ gauge theory (with Dirac fields) has been generalized. The
above observation implies that the super Lagrangian density for
the Dirac fields, using GIR (16) and super expansion (19), can be
written as
\begin{equation}
\tilde {\cal L}^{(d)} = \bar \Psi^{(G)} (x, \theta, \bar\theta)
\bigl (i\; \gamma^M\; D^{(h)}_M - m \bigr)\; \Psi^{(G)} (x, \theta, \bar\theta)
\equiv {\cal L}^{(d)},
\end{equation}
where $\gamma^M$'s are some non-trivial generalization of the $4
\times 4$ Dirac matrices $\gamma^\mu$ to the (4, 2)-dimensional
supermanifold and $\gamma^M D^{(h)}_M$ is defined as\footnote{The 
(4, 2)-dimensional representation
$\gamma^M \equiv (\gamma^\mu, C_\theta, C_{\bar\theta})$ serves our purpose in
equation (24). Here $C_\theta$ and $C_{\bar\theta}$ are  fermionic in nature
and they reduce to zero in the limit $(\theta, \bar\theta) \to 0$. The exact form
of $C_\theta$ and $C_{\bar\theta}$ is {\it not} essential for our purposes because,
irrespective of their form, we obtain $(\partial_\theta + i e \bar {\cal F}^{(h)})
\Psi^{(G)} = 0 $ and $(\partial_{\bar\theta} + i e {\cal F}^{(h)}) \Psi^{(G)} = 0$
when (24) is inserted into (23) for the verification of the equality (see, also, [13] for more details).}
\begin{equation}
\gamma^M D^{(h)}_M = \gamma^\mu \;(\partial_\mu + i e {\cal
B}^{(h)}_\mu) + C_\theta\;(\partial_\theta + i e \bar {\cal F}^{(h)}) +
C_{\bar \theta}\;(\partial_{\bar\theta} + i e {\cal F}^{(h)}).
\end{equation}
Here the superfields ${\cal B}_\mu^{(h)}, {\cal F}^{(h)}$ and
$\bar {\cal F}^{(h)}$ are the expanded form of equation (9) that
have been obtained after the application of HC.

It is straightforward to capture now the (anti-)BRST invariance
(i.e. $s_{(a)b}\; [\bar\psi (i \gamma^\mu D_\mu - m) \psi] = 0$)
of the Dirac part of the Lagrangian density (1) in the language of the superfields.
This can be expressed as follows
\begin{eqnarray}
s_b {\cal L}^{(d)} &=& 0 \;\;\Leftrightarrow\;\; \mbox{Lim}_{\theta \to 0}
\frac{\partial}{\partial \bar\theta} \tilde {\cal L}^{(d)} = 0, \nonumber\\
s_{ab} {\cal L}^{(d)} &=& 0 \;\;\Leftrightarrow\;\; \mbox{Lim}_{\bar \theta \to 0}
\frac{\partial}{\partial \theta} \tilde {\cal L}^{(d)} = 0.
\end{eqnarray}

\section{Nilpotent and anticommuting (anti-)BRST charges: superfield formulation}

It is straightforward to check that the total Lagrangian density (1)
transforms, under the (anti-)BRST symmetry transformations (cf.(2),(3)), as
\begin{eqnarray}
s_{ab} {\cal L}_b = \partial_\mu [B \partial^\mu \bar C],\; \qquad
s_{b} {\cal L}_b = \partial_\mu [B \partial^\mu C].
\end{eqnarray}
The Noether (anti-)BRST conserved currents $J^\mu_{(a)b}$
(i.e. $\partial_\mu J^\mu_{i} = 0, i = b, ab$),
that ensue due to the above symmetry transformations, are
\begin{eqnarray}
J^\mu_{ab} &=& B \partial^\mu \bar C - F^{\mu\nu} \partial_\nu \bar C - e \bar \psi \gamma^\mu \bar C \psi,
 \nonumber\\
J^\mu_{b} &=& B \partial^\mu  C - F^{\mu\nu} \partial_\nu  C - e \bar \psi \gamma^\mu  C \psi,
\end{eqnarray}
which lead to the derivation of the following conserved charges
\begin{eqnarray}
Q_{ab} = {\displaystyle \int}\; d^3 x \;\bigl [ B \dot {\bar C} - \dot B \bar C \bigr ], \qquad
\;\;Q_{b} = {\displaystyle \int} \;d^3 x \;\bigl [ B \dot { C} - \dot B  C \bigr ].
\end{eqnarray}
In the derivation of the above simple expressions 
for $Q_{(a)b}$ as well as in the proof of
the conservation of currents, the following equations of motion
\begin{eqnarray}
&&\partial_\mu F^{\mu\nu} = \partial^\nu B + e \bar \psi \gamma^\nu \psi,\; \qquad 
\Box C = \Box \bar C = \Box B = 0, \nonumber\\
&& i \gamma^\mu (\partial_\mu \psi) = m \psi + e \gamma^\mu A_\mu \psi, \qquad
i  (\partial_\mu \bar \psi)   \gamma^\mu = - m \bar \psi - e  \bar \psi \gamma^\mu A_\mu, 
\end{eqnarray}
that emerge from the Lagrangian density (1), have been exploited.

To understand the nilpotency and anticommutativity of the (anti-)BRST
charges, we have to express them in terms of the superfields that have been
obtained after the application of the HC and GIR (cf.(9),(19)). For instance, it can be
checked that the (anti-)BRST charges can be written as
\begin{eqnarray}
Q_b = i {\displaystyle \frac{\partial}{\partial \bar\theta} \frac{\partial}{\partial \theta}
\; \Bigl (\int}\; d^3 x \;  \bigl [ {\cal B}^{(h)}_0 \; F^{(h)} \bigr ] \Bigr ), \quad
Q_{ab} = i {\displaystyle \frac{\partial}{\partial \bar\theta} \frac{\partial}{\partial \theta}
\; \Bigl (\int}\; d^3 x \; \bigl [ {\cal B}^{(h)}_0 \; \bar F^{(h)} \bigr ] \Bigr ).
\end{eqnarray}
The above expressions immediately imply (cf.(12)-(15))
\begin{eqnarray}
{\displaystyle \frac{\partial}{\partial \bar\theta}} \; Q_b = 0,\; \qquad
{\displaystyle \frac{\partial}{\partial \theta}} \; Q_b = 0,\; \qquad
{\displaystyle \frac{\partial}{\partial \bar\theta}} \; Q_{ab} = 0,\; \qquad
{\displaystyle \frac{\partial}{\partial \bar\theta}} \; Q_{ab} = 0,
\end{eqnarray}
because of the fact that $ \partial_\theta \partial_{\bar\theta} + \partial_{\bar\theta}
\partial_\theta = 0$ and $\partial_\theta^2 = \partial_{\bar\theta}^2 = 0$. In the 
language of the (anti-)BRST symmetry transformations (cf.(10),(20)), we have 
\begin{eqnarray}
s_b Q_b = 0,\; \qquad s_{ab} Q_b = 0,\; \qquad s_b Q_{ab} = 0,\; \qquad s_{ab} Q_{ab} = 0,
\end{eqnarray}
which imply the nilpotency (i.e. $Q_{(a)b}^2 = 0$) 
and anticommutativity (i.e. $Q_b Q_{ab} + Q_{ab} Q_b = 0$)
properties of the conserved (anti-)BRST charges
because $s_b Q_b = - i \{Q_b, Q_b \} = 0, s_{ab} Q_b = - i \{Q_b, Q_{ab} \} = 0,
s_{ab} Q_{ab} = - i \{Q_{ab}, Q_{ab} \} = 0, s_{b} Q_{ab} = - i \{Q_{ab}, Q_{b} \} = 0$, etc.
It is interesting to point out that the (anti-)BRST charges (30) can be equivalently
expressed as
\begin{eqnarray}
Q_{ab} = i {\displaystyle \int d^3 x \int d^2 \theta}\; \Bigl [ {\cal B}^{(h)}_0 \; \bar F^{(h)} \Bigr ],
\quad
Q_{b} = i {\displaystyle \int d^3 x \int d^2 \theta}\; \Bigl [ {\cal B}^{(h)}_0 \; F^{(h)} \Bigr ],
\end{eqnarray}
where we have adopted $d^2 \theta = d \bar \theta d \theta$. Ultimately, the above expressions
imply the following interesting relationships
\begin{eqnarray}
i s_b s_{ab}\; \bigl [A_0 \bar C \bigr ] = B \dot {\bar C} - \dot B \bar C, \qquad\;\;\;
i s_b s_{ab} \;\bigl [A_0  C \bigr ] = B \dot {C} - \dot B  C,
\end{eqnarray} 
which establish the (anti-)BRST invariance of the (anti-)BRST charges.

The BRST charge $Q_b$, expressed in (30), can be also written in the following 
two distinctly different ways, namely;
\begin{eqnarray}
Q_b &=& i \; \mbox{Lim}_{\theta \to 0} {\displaystyle \frac{\partial}{\partial \bar\theta}\; \int}
\;d^3 x \; \bigl [ \dot {\bar F}^{(h)} F^{(h)} - i {\cal B}_0^{(h)} B(x) \bigr ] \nonumber\\
&\equiv& i {\displaystyle \int}\; d^3 x {\displaystyle \int}\;
d \bar \theta \;\bigl [ \dot {\bar F}^{(h)} F^{(h)}
 - i {\cal B}_0^{(h)} B(x) \bigr ], \nonumber\\
Q_b &=& - i \; \mbox{Lim}_{\bar\theta \to 0}  {\displaystyle \frac{\partial}{\partial \theta}\; \int}
\;d^3 x \;\bigl [ \dot {F}^{(h)} F^{(h)} \bigr ]
\equiv - i {\displaystyle \int}\; d^3 x {\displaystyle \int}
d  \theta \;\bigl [ \dot { F}^{(h)} F^{(h)} \bigr ].
\end{eqnarray}
 The above expressions demonstrate the followings

 (i) the nilpotency property because $\partial_\theta^2 = \partial_{\bar\theta}^2 =0$,

 (ii) the (anti-)BRST invariance of the BRST charge $Q_b$ because of the validity of (31) 
as well as the sanctity of the following expressions
\begin{eqnarray}
&& Q_b = i \; {\displaystyle \int} d^3 x \;  s_b \;\bigl [ \dot {\bar C} C - i A_0 B \bigr ]
\;\;\Rightarrow \;\;s_b Q_b = 0, \nonumber\\ 
&& Q_b = - i {\displaystyle \int} d^3 x \;  s_{ab} \;\bigl [ \dot {C} C \bigr ]
\;\;\;\;\Rightarrow \;\;\;\;\;s_{ab} Q_b = 0, 
\end{eqnarray}

  (iii) the anticommutativity property because $s_{ab} Q_b = - i \{Q_b, Q_{ab} \} = 0$.
  
\noindent
Thus, we note that the nilpotency and anticommutativity properties of the BRST charge
becomes quite simple in the language of the superfield formulation when QED
is considered on the (4, 2)-dimensional supermanifold.

In exactly similar fashion, we can express the anti-BRST charge as
\begin{eqnarray}
Q_{ab} &=& - i \; \mbox{Lim}_{\bar \theta \to 0} {\displaystyle \frac{\partial}{\partial \theta}\; \int}
\;d^3 x \; \bigl [ \dot {F}^{(h)} \bar F^{(h)} + i {\cal B}_0^{(h)} B(x) \bigr ] \nonumber\\
&\equiv& - i {\displaystyle \int}\; d^3 x {\displaystyle \int}\;
d  \theta \;\bigl [ \dot {F}^{(h)} \bar F^{(h)}
 + i {\cal B}_0^{(h)} B(x) \bigr ], \nonumber\\
Q_{ab} &=&  i \; \mbox{Lim}_{\theta \to 0}  {\displaystyle \frac{\partial}{\partial \bar \theta}\; \int}
\;d^3 x \;\bigl [ \dot {\bar F}^{(h)} \bar F^{(h)} \bigr ]
\equiv  i {\displaystyle \int}\; d^3 x {\displaystyle \int}
d  \bar \theta \;\bigl [ \dot {\bar F}^{(h)} \bar F^{(h)} \bigr ].
\end{eqnarray}
The above expressions automatically imply the validity of relations quoted in (31). In other words,
the (anti-)BRST charges (and their corresponding symmetry transformations) are always found
to be absolutely anticommuting and nilpotent of order two within the framework
of our present superfield approach to BRST formalism (as far as QED with Dirac fields is concerned).\\

\section{Conclusions}

In our present endeavor, we have concentrated on the (anti-)BRST
invariance of the Lagrangian density of a 4D interacting U(1)
gauge theory with Dirac fields. As in our earlier work [18] on the
4D {\it free} (non-)Abelian 1-form gauge theories (having no interaction with matter fields), 
we find that the
Grassmannian independence of the super Lagrangian densities (that
are expressed in terms of the superfields obtained after the
application of the HC and GIR) is a sure guarantee that the
corresponding 4D Lagrangian density would respect the nilpotent
(anti-)BRST symmetry invariance.

In the language of the geometry on the (4, 2)-dimensional
supermanifold, if the translation
of the super Lagrangian densities along

(i) the $\bar\theta$-direction of the above supermanifold (without
any translation along the $\theta$-direction) is zero, the
corresponding 4D Lagrangian density would possess the nilpotent
BRST invariance,

(ii) the $\theta$-direction of the above supermanifold (without
any shift along the $\bar\theta$-direction) is zero, there will be
nilpotent anti-BRST invariance for the 4D Lagrangian density of
the theory, and

(iii) the $\theta$- and $\bar\theta$-directions (one followed by
the other; either ways) is zero, there would be existence of the
(anti-)BRST symmetry invariance {\it together} for the 4D
Lagrangian density of the theory.

We have been able to show in our present work (as well as in our
earlier works [10-18]) that the nilpotent internal (anti-)BRST
symmetry transformations $s_{(a)b}$ for the 4D theories are very
intimately connected with the translational generators
($\partial_\theta, \partial_{\bar\theta}$) along the Grassmannian
directions of the (4, 2)-dimensional supermanifold. Thus, one of
the key features of our superfield approach to BRST formalism is
the sure guarantee that the nilpotency (i.e. $s_{(a)b}^2 = 0$) and
the anticommutativity (ie. $s_b s_{ab} + s_{ab} s_b = 0$) properties
would always be satisfied by the (anti-)BRST symmetry
transformations $s_{(a)b}$.

We have already
noted that, the above specific features are the integral ingredients of our
superfield approach to BRST formalism because the translational
generators ($\partial_\theta, \partial_{\bar\theta}$) always obey the nilpotency
property ($\partial_{\theta}^2 = 0, \partial_{\bar\theta}^2 = 0$) as well as the 
anticommutativity property (i.e. $\partial_\theta
\partial_{\bar\theta} + \partial_{\bar\theta} \partial_\theta =
0$). We have been able to shed more light on it through the (anti-)BRST
charges when we have expressed them in terms of the superfields
and translational generators (see, section 5). The anticommutativity
and nilpotency ensue automatically.

The important geometrical consequences of the these GIRs on
the matter superfields (cf. (16) and (38) below) are as follows

(i) the Grassmannian independence of the Dirac part of the super
Lagrangian density ($\tilde {\cal L}^{(d)}$) due to the
application of (16), and

(ii) the Grassmannian independence of the kinetic energy term for
the U(1) gauge superfield and the super Lagrangian density $\tilde
{\cal L}^{(d)}$ due to the application of the GIR (38) (see, Appendix below).

In fact, the GIR (38) (see, Appendix below)
on the matter superfields provides a generalization of the HC
because it leads to results that are obtained due to the
application of HC and GIR (16) separately.

One of the highlights of our present investigation is the
simplicity and beauty that have been brought in for the
(anti-)BRST invariance of the Lagrangian density of the 4D
interacting U(1) gauge theory within the framework of the
superfield approach to BRST formalism. It is nice to point out that
our present work has {\it already} been generalized to 

(i)  the case of interacting Abelian U(1)
gauge theory where there is coupling between the gauge field 
and complex scalar fields [19], and 

(ii) the case of the interacting 4D
non-Abelian gauge theory (with Dirac fields) which happens to be
more general than our present theory [20].

We have also devoted time on the nilpotent
(anti-)BRST as well as the nilpotent (anti-)co-BRST invariance of
the 4D free Abelian 2-form gauge theory in the Lagrangian
formalism [21]. It would be interesting endeavor to capture the
nilpotent (anti-)BRST (as well as (anti-)co-BRST) invariance of
the 4D (non-)Abelian 2-form (and still higher form gauge theories) 
within the framework of the superfield approach to BRST formalism. These are
some of the issues that are presently under investigation.\\

\noindent
{\bf Acknowledgements:} A part of the present work was carried out at SISSA.
It is a pleasure to thank the Director (SISSA, Trieste, Italy)
for the very warm hospitality extended to me. Thanks are also due to L. Bonora 
for invitation and some fruitful discussions. Financial support from the
Department of Science and Technology (DST), Government of India,  
under the SERC project sanction grant No: - SR/S2/HEP-23/2006, is gratefully acknowledged, too.  \\

\noindent
{\large {\bf Appendix A: on (anti-)BRST transformations for all the fields
from a single GIR on matter superfield}}\\

In this Appendix, we concisely recapitulate some of the key points connected with
the derivation of the results of sections 3 and 4 from a single
GIR on the superfields, defined on the (4, 2)-dimensional
supermanifold [18]. This GIR also owes its origin to the (super)
covariant derivatives but, in a form, that is quite different from
(16). The explicit form  of this GIR is
\begin{equation}
\bar \Psi (x,\theta,\bar\theta) \; \tilde D \; \tilde D\; \Psi
(x,\theta,\bar\theta) = \bar \psi (x) \; D\; D\; \psi (x),
\end{equation}
where $\tilde D$ and $D$ are the covariant derivatives defined on
the (4, 2)-dimensional supermanifold and 4D flat Minkowski
spacetime manifold, respectively\footnote{ The stunning
strength of this equation does not appear in the context of the
Abelian U(1) gauge theory because all the fields are commutative
in nature. Its immense mathematical power gets reflected in its
full blaze of glory in the non-Abelian gauge theory where it leads
to the {\it exact} derivation of all the nilpotent (anti-)BRST
symmetry transformations for {\it all} the fields of the 4D non-Abelian
gauge theory [13].}. These are defined as follows
\begin{equation}
\tilde D = \tilde d + i e \tilde A^{(1)}, \qquad D = d + i e A^{(1)},
\qquad d = dx^\mu \partial_\mu, \qquad A^{(1)} = dx^\mu A_\mu,
\end{equation}
where all the quantities, in the above, have been taken from the
earlier sections 3 and 4. For instance, equations (5), (7) and
(17) have been used.

The above restriction is a GIR on the superfields (defined on the
above supermanifold) because of the fact that the r.h.s. of (38)
is
\begin{equation}
\bar\psi D D \psi = \frac{ie}{2} (dx^\mu \wedge dx^\nu) \bar\psi
F_{\mu\nu} \psi \equiv i e \bar\psi F^{(2)} \psi.
\end{equation}
It is straightforward to check that the above quantity is U(1)
gauge invariant. A noteworthy point, at this stage, is that the
r.h.s. is a 2-form with the differentials (i.e. $dx^\mu \wedge
dx^\nu)$ in the spacetime variables {\it only}. However, the
l.h.s. of the equation (26) contains all the differential 2-forms
in terms of the superspace variables. It is obvious that all the
coefficients of the differentials in the Grassmannian variables of
the l.h.s. will be set equal to zero.

It has been clearly demonstrated in our earlier work [18] that the
outcomes of the above equality in (38) (i.e. GIR) are the
relationships that we have already obtained separately and
independently in (8) and (18). Thus, the expansions of the
superfields, ultimately, reduce to the forms which can be
expressed in terms of the appropriate
(anti-)BRST symmetry transformations
$s_{(a)b}$ as quoted in the key equations (9) and (19).

It is obvious, from our above discussions, that the total
(anti-)BRST invariant Lagrangian density (1), defined in terms of
the local fields (taking their values on the 4D flat spacetime
manifold), can be recast in terms of the superfields (defined on
the (4, 2)-dimensional supermanifold) by adding super Lagrangian
densities given in equations (11) and (23) as
\begin{equation}
\tilde {\cal L}_T = \tilde {\cal L}_{(g)}^{(1,2,3)} + \tilde {\cal L}^{(d)}.
\end{equation}
Now the nilpotent (anti-)BRST invariance of the Lagrangian density
(1) can be expressed as (25) with the
replacement: $\tilde {\cal L}^{(d)} \to \tilde {\cal L}_{T}$.
Finally, we conclude that the Grassmannian independence of the
super Lagrangian density encodes the (anti-)BRST invariance of the
4D Lagrangian density.

\end{document}